\documentclass[aps,prl, twocolumn, showpacs, superscriptaddress,preprintnumbers, amsmath, epsfig, floatfix,normalem]{revtex4-1}

\usepackage{graphicx}
\usepackage{physics}
\usepackage{dcolumn}
\usepackage{bm}
\usepackage{ulem}
\usepackage{subfigure}
\usepackage[utf8]{inputenc}
\usepackage[english]{babel}
\usepackage[usenames,dvipsnames]{color}
\usepackage[]{ulem}
\usepackage[colorlinks=true,
linkcolor=blue,
filecolor=magenta,      
urlcolor=blue,
citecolor=blue]{hyperref}
\newcommand{\comment}[1]{\textcolor{red}{#1}}
\renewcommand{\comment}[1]{\relax}
\newcommand{\todelete}[1]{\textcolor{green}{\sout{#1}}}
\renewcommand{\todelete}[1]{\relax}

\newcommand*{\balancecolsandclearpage}{%
	\close@column@grid
	\clearpage
	\twocolumngrid}
	
\newcommand{\beginsupplement}{%
        \setcounter{table}{0}
        \renewcommand{\thetable}{S\arabic{table}}%
        \setcounter{figure}{0}
        \renewcommand{\thefigure}{S\arabic{figure}}%
     }

\begin{document}

\title{Classical linear chain behavior from dipolar droplets to supersolids}

\author{K. Mukherjee}
\author{S.M. Reimann}
\affiliation{Mathematical Physics and NanoLund, Lund University, Box 118, 22100 Lund, Sweden}

\begin{abstract}
We investigate the classicality of linear dipolar droplet arrays through a normal mode analysis of the dynamical  properties in comparison to the supersolid regime. The vibrational patterns of isolated-droplet crystals that time-evolve 
after a small initial kick closely follow the properties of a linear droplet chain. For larger kick velocities, however, droplets may coalesce and separate again, showing distinct deviations from classicality.  In the  supersolid regime the normal modes are eliminated by a counter-flow of mass between the droplets, signaled by a reduction of the center-of-mass motion. 
\end{abstract}
%\pacs{47.20.Dr, 47.35.Pq, 47.54.-r}
\date{\today}

\maketitle

In dipolar Bose-Einstein condensates (dBEC) intriguing new quantum phases of matter at ultra-low temperatures were discovered (see e.g. the reviews~\cite{Baranov2008,Lahaye2009,Baranov2012,Boettcher2021,Chomaz2022}). 
Remarkably, a trapped dBEC can develop a spontaneous periodic density modulation while maintaining coherence and frictionless flow, resulting in a dipolar supersolid~\cite{Boettcher2019,Tanzi2019a,Chomaz2019,Natale2019,Tanzi2019b,Guo2019,Hertkorn2019,Hertkorn2021b}. 
Such state of matter with simultaneous off-diagonal and diagonal long-range order~\cite{Gross1957,*Gross1958,Yang1962,Andreev1969,Chester1970,Leggett1970,Pomeau1994,Boninsegni2012} has long been debated for $^4$He~\cite{Kim2004a,*Kim2004b,Balibar2010,Kim2012,Boninsegni2012,Chan2013}, followed by alternative setups with ultra-cold atoms ~\cite{Henkel2010,Cinti2010,Saccani2011,Leonard2017a,*Leonard2017b,Lin2011,Li2016,Li2017}.  
In dBECs, the  mechanism driving the formation of broken-symmetry states originates from the interplay between inter-particle interactions and quantum fluctuations~\cite{Lima2011,Wachtler2016a,*Wachtler2016b,Bisset2016}.  In experiments with dysprosium~\cite{Kadau2016,Ferrier2016,Schmitt2016,Boettcher2019b} or erbium~\cite{Chomaz2016,Chomaz2018} it was found that density modulations and self-bound filaments stabilize  similarly to droplets of binary Bose gases~\cite{Petrov2015, Petrov2016, Arlt2018} realized with potassium in different hyperfine states~\cite{Cabrera2017, Semeghini2018, Skov2021}. 
The dBEC includes both short- and long-range interactions, and by increasing the relative strength of the long-range interaction, there is a transition from a superfluid to a supersolid and a crystal phase where the droplets become almost isolated from one another~\cite{Wachtler2016a,*Wachtler2016b,Roccuzzo2019}. 
 In addition to the above references there is a large volume of works addressing dipolar droplets  and supersolids, see for example 
\cite{Bisset2015,Xi2016,Blakie2016,Saito2016,Bisset2016,Macia2016,Baillie2017,Edler2017,Baillie2018,Roccuzzo2019,Zhang2019,
Blakie2020, Roccuzzo2020, Gallemi2020, Mishra2020, Chomaz2020,Poli2021,Hertkorn2021,Hertkorn2021b, Zhang2021,Bland2022,Young2022, Ghosh2022,Gallemi2022,Schmidt2022,Halder2022,Tengstrand2021, Roccuzzo2022,Sindik2022}. Dipolar mixtures~\cite{Smith2021,Bisset2021,Li2022, Scheiermann2022,Bland2022b, Halder2022b} were also discussed.
Recent experimental efforts are directed towards two-dimensional systems, see e.g., \cite{Norcia2021,Schmidt2021,Biagioni2022,Bland2022} 
and vorticity~\cite{Klaus2022}, and to explore the out-of-equilibrium dynamics~\cite{Tanzi2019b,Sohmen2021,Ilzhofer2021,Biagioni2022,Norcia2022}.

Although the underlying origin of the self-bound droplets is purely quantum mechanical, 
there is an aspect of \textit{classicality}: The droplets may localize and rigidly organize themselves in lattice-periodic structures, 
but may also phase-coherently overlap forming a supersolid. An intriguing and unresolved question is, to what extent classicality may prevail in these systems. 
 \begin{figure}
 	\centering
 	\includegraphics[width = 0.5\textwidth]{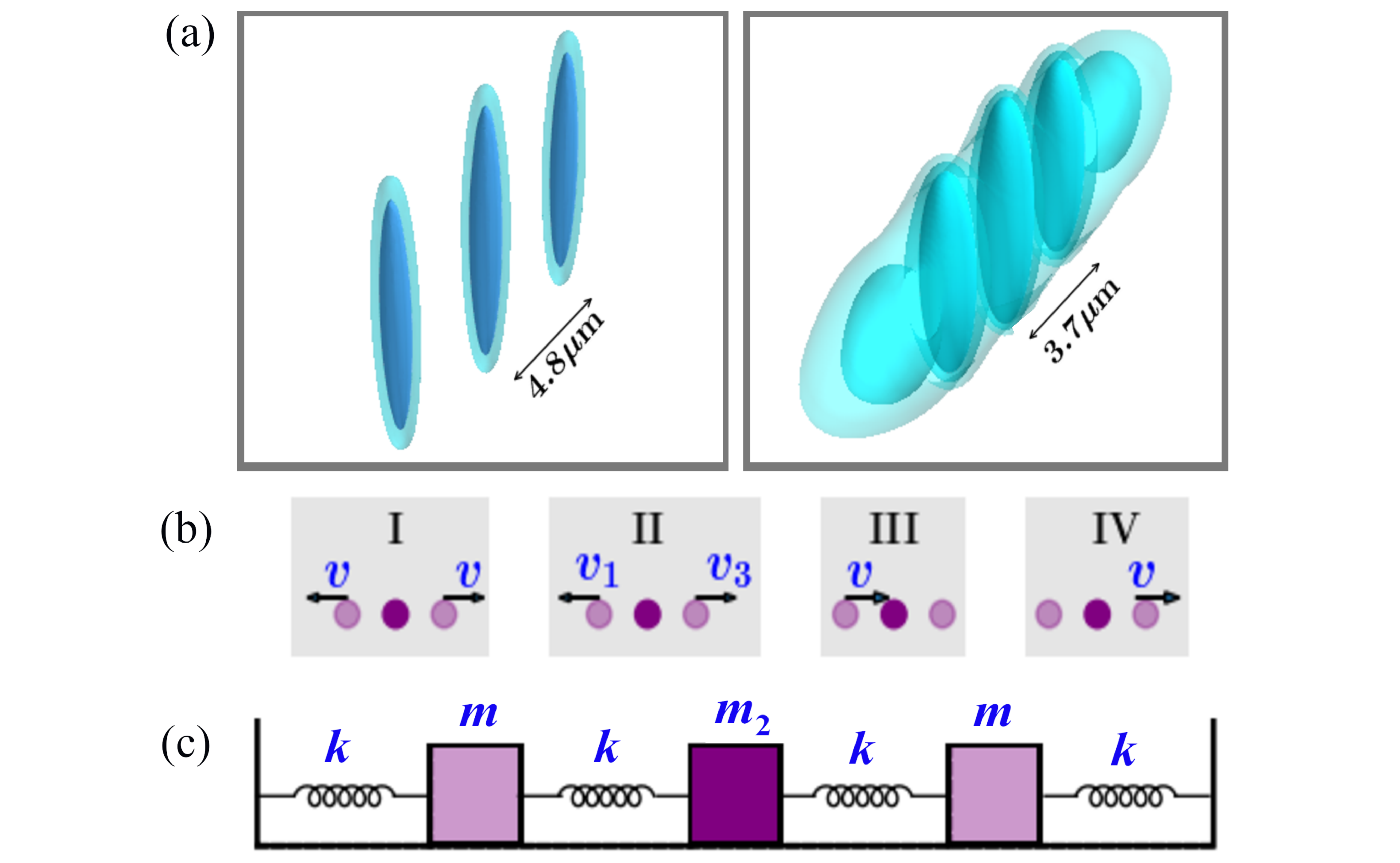}
 	\caption{(Color online) (a) Iso-surfaces of three-dimensional density depicting isolated droplets {\it (left)} and a supersolid phase {\it (right)}, realized by a system of $N =35,000$ $^{164}$Dy atoms at scattering length $a_{s}=84a_{0}$ and $a_{s}=94a_{0}$, respectively. (b) Sketch of different configurations where one or two droplets are kicked with a velocity. (c) Sketch of the linear chain model of the linear droplet array, with masses $m < m_2$ attached by springs of stiffness $k$.}
 	\label{Fig0}
 \end{figure}
In this Letter, we demonstrate a mechanism to gauge the classicality of these novel supersolid or droplet array states by their 
dynamics, based on a classical linear-chain model~\cite{Rutherford, Pippard} in comparison to a dynamical Gross-Pitaevskii (eGP) approach. 
We consider a few droplets in a dBEC confined by a three-dimensional elongated harmonic trap (see Fig.~\ref{Fig0}(a)),
similar to the recent experiments in~\cite{Boettcher2019, Tanzi2019a,Hertkorn2019,Chomaz2019,Hertkorn2021b}. 
To initiate the vibrational dynamics, we give one or two droplets a kick along the weakly confined direction, as sketched 
in Fig.~\ref{Fig0}(b). Remarkably, a classical spring-mass model as sketched in Fig.~\ref{Fig0}(c) can perfectly account for the vibrational patterns, provided the droplets are fully separated from each other.  The observed vibrational frequencies hint at the triggering of underlying collective excitation modes~\cite{Hertkorn2019,Hertkorn2021b}. A finite background density  interconnecting the droplets however destroys the resemblance to classical patterns by inducing particle flow between the droplets. When the kick velocity is large enough compared to the velocity scale imposed by the harmonic confinement, droplets that are initially isolated  can collide, transferring particles from outer to central droplets, a phenomenon that mimics the collision of classical liquid droplets~\cite{ashgriz1990, qian1997}. This collision destroys the above-mentioned  vibrational pattern as well. 
For binary quantum droplets with short-range interactions~\cite{Cabrera2017, Semeghini2018} 
collision dynamics previously has been reported in experiment~\cite{Ferioli2019} and theory~\cite{Cikojevic2021,Alba2022}. 
In contrast, in the present work the collision dynamics between droplets is realized within a single confined species. Our findings offer a new avenue to assess classicality and the dynamical manifestation of collective excitations in droplet arrays at the verge between the crystal and supersolid phase.  
\par
{\it \textcolor{blue}{Model}.--} We consider a dBEC of atoms with mass $M$ and magnetic dipole moment $\mu$ aligned along the $z$ axis, harmonically trapped  by $V(r) = M(\omega^2_x x^2 + \omega^2_y y^2 + \omega^2_z z^2)/2$. The frequencies $\omega_i$ (where $ i=x,y,z$) satisfy $\omega_x < \omega_y, \omega_z$ resulting in an elongated geometry along $x$. At  zero  temperature, the  system  is well described by the eGP equation 
~\cite{Lima2011,Wachtler2016a,*Wachtler2016b, Chomaz2016}
\begin{eqnarray}\label{eGPE}   
& i\hbar \frac{\partial \psi(\textbf{r},t)}{\partial t}  =  \bigg[-\frac{\hbar^2}{2M}\nabla^2 + V(\textbf{r}) + g \abs{\psi(\textbf{r},t)}^2+\nonumber\\& \gamma(\epsilon_{dd})\abs{\psi(\textbf{r},t)}^3 +\int dr^{\prime} U_{dd}(\textbf{r-r}^{\prime})\abs{\psi(\textbf{r}^{\prime},t)}^2 \bigg] \psi(\vb{r},t), 
\end{eqnarray}
where $g = 4 \pi \hbar^2a_s/M$ is the short-range repulsive contact interaction fixed by the scattering length $a_s$, and the dipolar interaction $U_{dd} (\textbf{r},t) = \frac{\mu_0 \mu^2_{m}}{4\pi} \left[\frac{1-3\cos^2\theta}{\vb{r}^3}\right]$ with $\theta$ being the angle between $\vb{r}$ and the $z$-axis. The final term in Eq.~\eqref{eGPE} is given by the repulsive Lee-Huang-Yang (LHY) correction 
with $\gamma(\epsilon_{dd}) = \frac{32}{3}g \sqrt{\frac{a_s^3}{\pi}} \left(1+\frac{3}{2}\epsilon_{dd}^2\right)$~\cite{Lima2011, Lima2012}. The dimensionless parameter $\epsilon_{dd} = a_{dd}/a_s$ with dipolar length $a_{dd}= \mu_0 \mu^2_{m} M /12\pi\hbar^2$ quantifies the relative strength of the DDI as compared to the contact interaction, and $v_{\rm osc} = \sqrt{\hbar \omega_x/M}$ sets the characteristic velocity scale. Equation~\eqref{eGPE} is
solved using a split-step Crank-Nicholson method~\cite{crank_nicolson1947,Antoine2013,supmat}
in imaginary time to obtain the initial ground state and in real time to monitor the dynamics (see Supplemental Material~\cite{supmat} for details).
The system showcases a superfluid phase for a sufficiently small value of $\epsilon_{dd}$. Increasing this parameter, the supersolid phase (SS) is favored in a  window of values of $\epsilon_{dd}$ beyond which one enters the isolated droplet phase ($\rm DL_{I}$). \par 
In the following, we utilize the experimentally relevant~\cite{Tanzi2019a, Boettcher2019} parameters of $^{164}$Dy dipolar BEC, namely, $\omega_x/(2\pi) = 19 \rm Hz$,          $\omega_y/(2\pi)= 53 \rm Hz$, and $\omega_z/(2 \pi) = 87 \rm Hz$, and $N = 35,000$. Modulated density profiles are found for $a_s < 94.9a_{0}$. The density isosurfaces  of the $\rm  DL_{I}$ and $\rm SS$ phases, realized at $a_s = 84a_{0}$ and $94a_{0}$, respectively, are shown in Fig~\ref{Fig0}(a). Having determined the ground state, showing the three localized density structures (droplets), we proceed to investigate the vibrational and collisional dynamics of these dipolar droplets. In order to accomplish this, we give a specific droplet an initial kick with velocity $v$ at $t=0$ in the following manner:
Firstly, the central-position ($x_{0}$) of a droplet along the  $x$-axis in SS and $\rm DL_{I}$ phases is determined by locating the peaks in the ground state density profiles. Then we apply a quench through a transformation of the order parameter, 
\begin{equation}
	\psi(x, y,z) = \psi(x, y, z) e^{iv\mathcal{F}(x)x}.
\end{equation}
%\[
%\psi(x,y,z)= 
%\begin{cases}
%\psi(x,y,z)e^{ivx},& %\abs{(x-x_{0})}\leq L/2\\
%\psi(x,y,z).              & %\text{otherwise}
%\end{cases}
%\]
Here, the function $\mathcal{F}(x) = A/(A + B\cosh(L(x-x_{0})))$, with $A$, $B$, and $L$ being constants, guarantees that each localized droplet has velocity $v$ across its spatial extent; beyond its extent the velocity falls to zero swiftly but continuously, maintaining the continuity of the wavefunction~(see Supplemental Material~\cite{supmat}). Having such kicked state
at hand, we evolve the system in real-time. Depending upon which droplets are being kicked and the initial conditions thereof, a variety of dynamical situations can be realized.  We consider the following four cases [see Fig.~\ref{Fig0}(b)]: Case \textbf{I} corresponds to kicking the left and right droplets with equal velocity in the $-x$ and $+x$ directions, respectively. In case \textbf{II}, the left and right droplets are kicked in the opposite direction with unequal velocity. Case \textbf{III} constitutes the left droplet being kicked in the $+x$ direction. Finally, the case \textbf{IV} deals with the scenario where only the central droplet is kicked in the $+x$ direction.
\begin{figure}
	\centering
	\includegraphics[width = 0.4\textwidth]{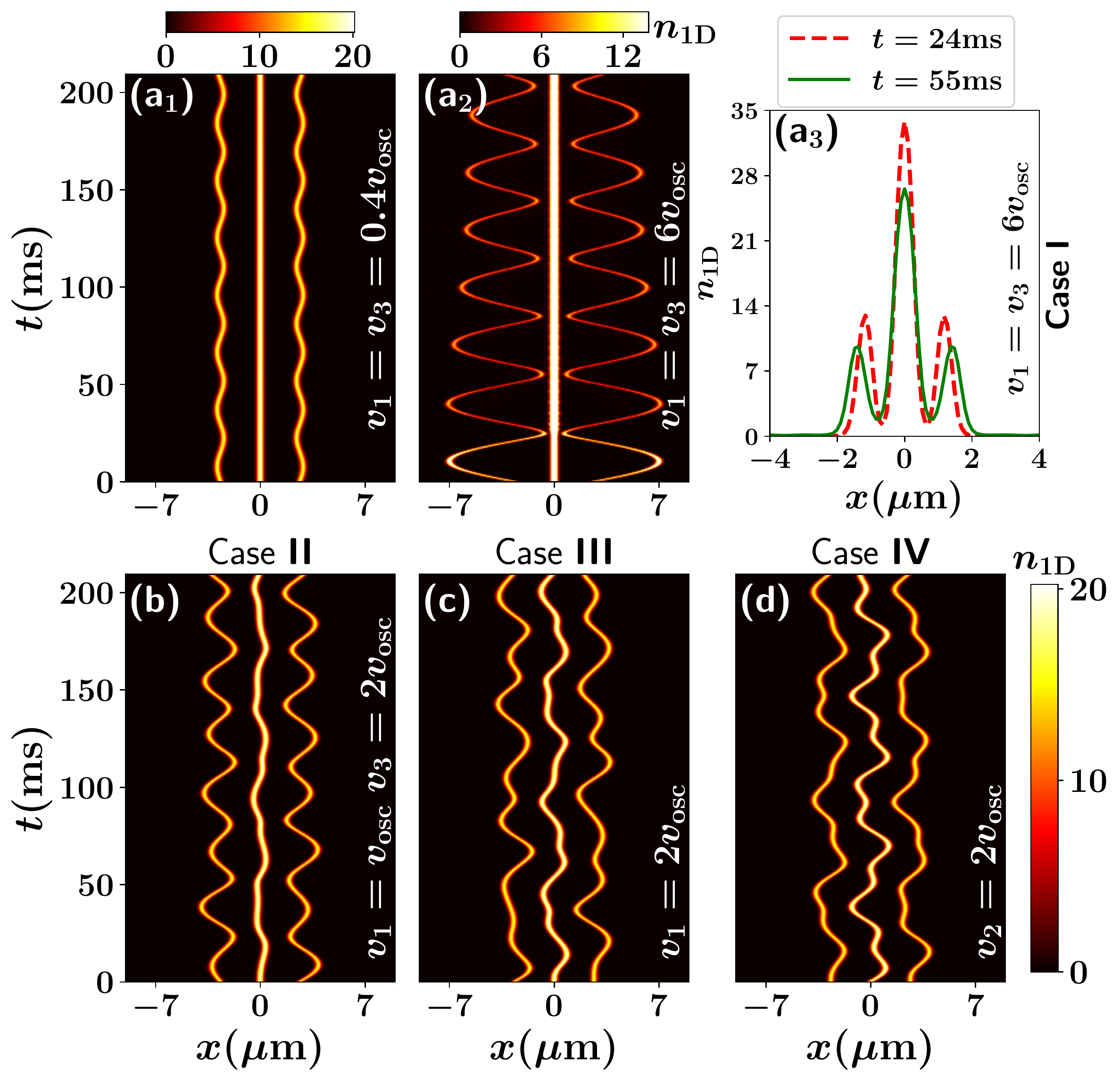}
	\caption{Time evolution of integrated density $n_{\rm 1D}$ of quasi-1D dipolar quantum droplets exhibiting (\text{a}$_1$) pure vibrational and (\text{a}$_2$) collisional dynamics for the case \textbf{I} when two outer droplets are kicked in the opposite direction with the same velocity (see the legends). (\text{a}$_3$) The integrated density profiles $n_{\rm 1D}$ during the first two collision times corresponding to (\text{a}$_2$). During the collision, the droplets make direct contact, exchanging particles from the outside to the central droplet [(\text{a}$_2$), (\text{a}$_3$)]. Shown also the vibrational dynamics within  (b) case \textbf{II}, (c) case \textbf{III} and (d) case \textbf{IV} that feature different initial velocity configurations(see also the legends). The colorbar represents 1D density in the unit of $1000\mu m^{-1}$. The colorbar in (a$_2$) has been adjusted to enhance the contrast. The $v_{\rm osc} = 2.1 \times 10^{-4} \rm m/s$ is the velocity scale.}
	\label{Fig1}
\end{figure}
\par
{\it \textcolor{blue}{Vibrational and collisional dynamics.}--} In order to elucidate the vibrational and collisional dynamics taking place along the weakly confined $x$-direction, 
we examine the time evolution of the one-dimensional (1D) integrated density profiles, $n_{\rm 1D}(x) = \int\abs{\psi(x,y,z)}^2 dy dz$, being experimentally detectable e.g. via \textit{in-situ} imaging~\cite{Hertkorn2021b,Chomaz2022}. Focusing first on the case \textbf{I} [Fig.~\ref{Fig1}(a$_{1}$)-(a$_{3}$)], we notice that the outer droplets exhibit periodic vibrational motion, while the center droplet stays motionless. For low enough velocity, $v < v_{\rm osc}$ [Fig.~\ref{Fig1}(a$_{1}$)], the motion is precisely of sinusoidal type with a $\pi$ phase difference between the trajectories of the left and right droplets. This resembles a fundamental mode of vibration of a corresponding  {\it classical} spring-mass system (see the discussion below).  
The amplitude of vibration increases for increasing velocity, $v > v_{\rm osc}$, and the two outer droplets can collide with the central one, resulting in particle exchange either from the outer droplets to the central one, or vice versa. For example, a particle transfer from the outer to central droplet upon collision can be noticed [Fig.~\ref{Fig1}(a$_2$)] when $v_1 =v_3 = 6 v_{\rm osc}$ . Note that a particle exchange occurs exclusively during the first two direct encounters between droplets, at $t \approx 24 \rm ms$ and $t \approx 55 \rm ms$, respectively. For $t > 55 \rm ms$, the amplitude of vibration of the outer droplets decreases, and as a result, they do not anymore come into direct contact with the central droplet, with no further collisions. To better resolve the collision, we also show in Fig.~\ref{Fig1}(a$_3$) the density profiles $n_{\rm 1D}$ at $t=24 \rm ms$ and $t = 55\rm ms$, respectively. Evidently, the droplets overlap during the collision, causing mass-flow. Consequently, the vibrational mode visible at low kick velocity (see, for example, Fig.~\ref{Fig1}(a$_2$)) is eliminated.     \par
Interestingly, for $v_1 \ne v_3$ (case \textbf{II}), the central droplet ceases to be stationary, see Fig.~\ref{Fig1}(b), signaling the onset of center of mass (COM) motion. Also, the trajectory of each droplet is not pure sinusoidal anymore. This  suggests that the emerging vibration is not caused just by a single fundamental mode but rather by a linear combination of different modes. A very similar dynamics is observed for the case \textbf{III}. The representative example illustrated in Fig.~\ref{Fig1}(c) indicates the involvement of multiple frequencies in the vibrational pattern. Now, the left droplet receives a kick along $+x$ direction, and it moves until it reaches the close vicinity of the central one at $t=7.5 \rm ms$. The central droplet then begins to migrate toward the right one, compelling the latter to move in the $+ x$ direction as well. Turning to the case \textbf{IV} [Fig.~\ref{Fig1}(d)], when the central droplet is kicked towards the right one, we notice that the outer droplets follow nearly identical trajectories, vibrating in phase, while the central one showcases out-of-phase vibration with the others, in contrast to the cases mentioned before. This drastic modification of the vibrational pattern arguably hints at the vanishing or softening of the fundamental mode responsible for the dynamics of  case \textbf{I},  setting the stage for 
analyzing the system in terms of its classical modes. 
\begin{figure}
	\centering
	\includegraphics[width = 0.4\textwidth]{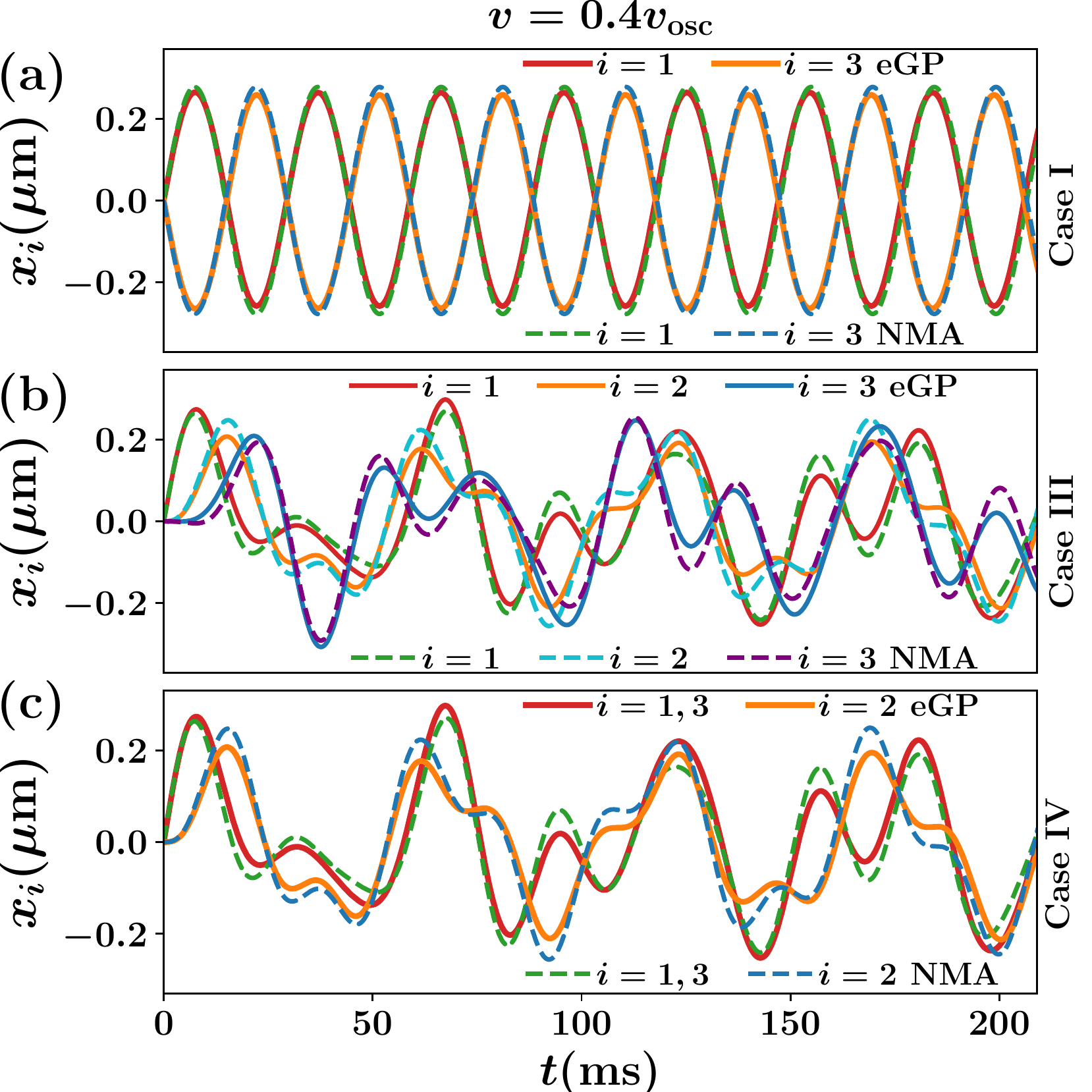}
	\caption{(Color online) Comparison  of droplet trajectories $x_{i}(t)$ between eGP simulation  and those obtained from the normal mode analysis(NMA) (see the legends) for $(\textbf{a})$ case \textbf{I}, $(\textbf{b})$ case \textbf{III} and $(\textbf{c})$ case \textbf{IV}, all for velocity $v = 0.4 v_{\rm osc}$. The NMA matches well with those obtained from the eGP for the characteristic frequencies $\omega_1/(2\pi)= 24\rm Hz$ and $\omega_2/(2 \pi) = 25.7 \rm Hz$. The eGP simulation is performed with $N=35,000$ particles having scattering length $a_{s}=84a_{0}$ and confined in a trapping potential with frequencies $\omega_x/(2 \pi) = 19\rm Hz$, $\omega_y/(2 \pi)= 53\rm Hz$, and $\omega_z/(2 \pi)= 87 \rm Hz$. }
\label{Fig2}
\end{figure}

{\it \textcolor{blue}{Normal modes of vibration.}--} In order to capture the fundamental modes of vibration responsible for the emergent oscillation pattern observed in the eGP simulation, we resort to the so-called normal mode analysis (NMA)~\cite{Rutherford}. In particular, we model the three harmonically localized droplets as three compact masses connected by springs with spring constant $k$, sketched in Fig.~\ref{Fig0}(c), assuming that the well-known Hooke's law~\cite{Pippard} is satisfied. Inspired by the eGP simulations, we assume that the central droplet has mass $m_2$, while the outer ones have equal masses $m_1=m_3=m$,  satisfying $m_2 > m$. Therefore, the two characteristic angular frequencies of the springs read $\omega_1 =\sqrt{k/m}$, and $\omega_2 = \sqrt{k/m_2}$. These frequencies are determined by the combined effect of confinement and interparticle interactions. The instantaneous configuration of the system is specified by the horizontal displacements of the three masses from their equilibrium positions, $\boldsymbol{X}(t) = (x_1(t), x_2(t), x_3(t))$. This is manifestly a three-degree of freedom system, and the three normal mode frequencies  (labeled from slow "$s$" to medium "$m$" to fast "$f$") are given by~(see Supplemental Material~\cite{supmat}) 
$\omega_s =  [\omega^2_1 + \omega^2_2 - \sqrt{\omega^{4}_1 + \omega^{4}_2}]^{1/2}$, 
$\omega_m =  \sqrt{2} \omega_1$, and
$\omega_f =  [\omega^2_1 + \omega^2_2 + \sqrt{\omega^{4}_1 + \omega^{4}_2}]^{1/2}$,  
with the associated normal modes being $\boldsymbol{A}_1 = (1, 0, -1)$, $\boldsymbol{A}_2 =(\omega^2_1, \omega^{2}_{-}, \omega^2_1)$ and $\boldsymbol{A}_3 =(\omega^2_1, \omega^{2}_{+}, \omega^2_1)$, respectively, where $\omega^2_{\pm} = (\omega^2_{2} - \omega^2_{1}) \pm \sqrt{\omega^{4}_1 + \omega^{4}_{2}}$. Note that the initial displacements of the masses are zero, namely, $x_1(t=0)=0$, $x_2(t=0)=0$, and $x_3(t=0)=0$. Consequently, the most general solution of the equation of motion, revealing trajectories of each individual mass, is given by $\boldsymbol{X}(t)=a_{m}\boldsymbol{A}_1 \sin(\omega_m t) + a_{f} \boldsymbol{A}_2 \sin(\omega_f t) + a_{s} \boldsymbol{A}_3 \sin(\omega_s t)$. The coefficients $a_m$, $a_f$, $a_s$ can be found from the three initial conditions on  the velocities.\par

\begin{figure}
	\centering
	\includegraphics[width = 0.4\textwidth, height = 0.46\textwidth]{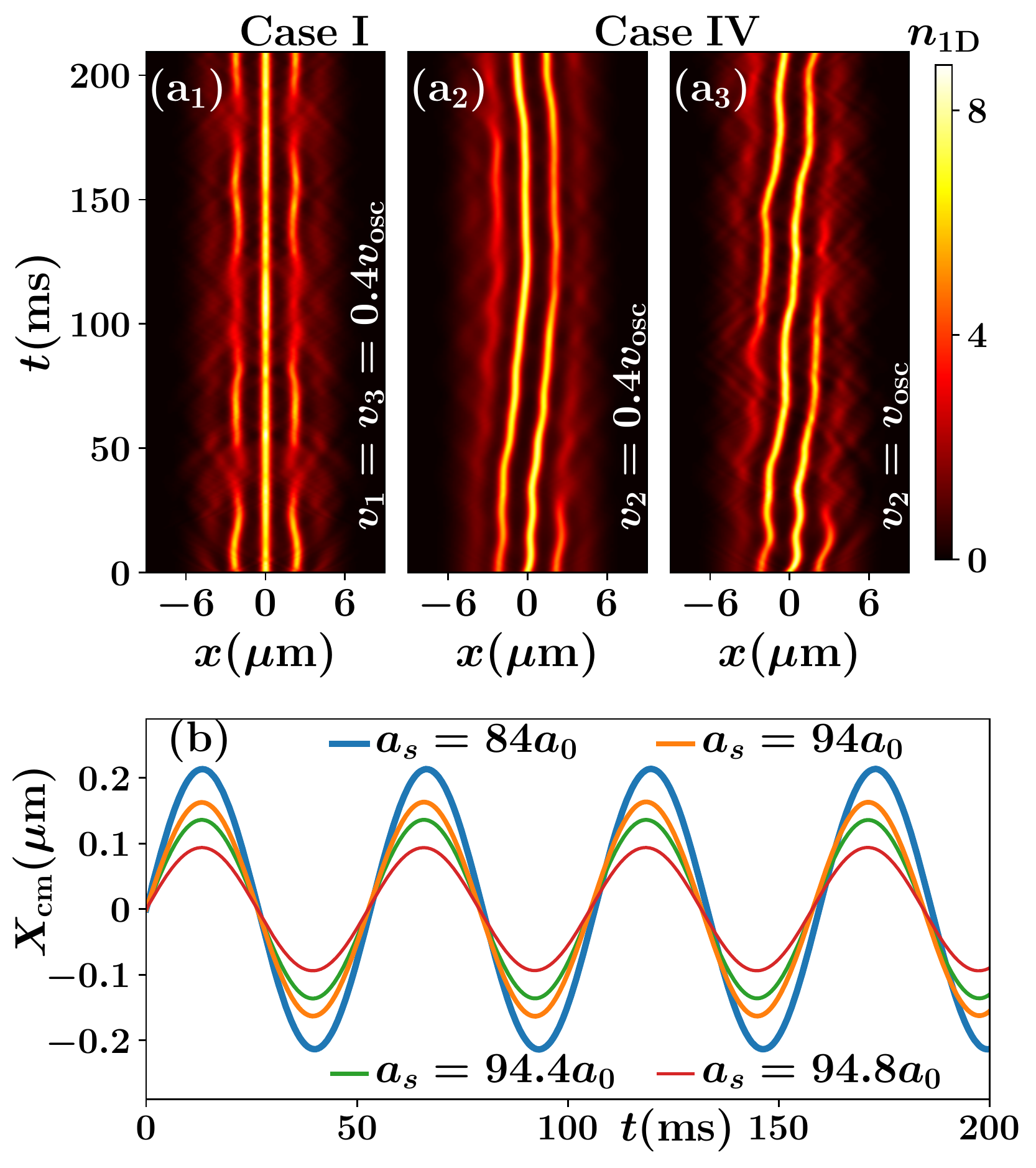}
	\caption{(Color online) Time evolution of integrated density $n_{\rm 1D}$ within the SS state focusing on (a$_1$) the case \textbf{I} and (a$_2$)-(a$_3$) the case \textbf{IV} for kick velocities $0.4 v_{\rm osc}$[(a$_1$), (a$_2$)] and $v_{\rm osc}$[(a$_3$)]. (b) The center-of-mass motion(for case \textbf{IV}) along the x-axis for different scattering lengths (see the legends) representing varying background density. The colorbar represents the 1D density in units of $1000\rm \mu m^{-1}$, and $v_{\rm osc}=2.1 \times 10^{-4}\rm m/s$  is the harmonic oscillator velocity scale}\label{Fig4}
\end{figure}

For the case \textbf{I}, the initial velocities read $\dot{x}_1(0)=-v$, $\dot{x}_2(0)=0$, $\dot{x}_3(0) = v$, which give the coefficients $a_f = a_s = 0$, and $a_m = v/(\sqrt{2} \omega_1)$. As a result, the trajectories can be calculated as $x_1(t)=a_{m}\sin(\omega_m t)$, $x_2(t) =0$, and $x_3(t) = - a_{m} \sin(\omega_m t)$. We also calculate the trajectories of each individual droplet $x_{j}(t)$, with $i=1,3$ from the eGP simulation for the case \textbf{I} (see Fig.~\ref{Fig2}(a)). Notably, for velocity $v < v_{\rm osc}$, the trajectories from the simulation match perfectly to those obtained from the NMA  with $\omega_1/(2 \pi) = 24 \rm Hz$, and $\omega_m /(2 \pi) = 33.94 \rm Hz $. Indeed, the dynamics in case \textbf{I} is governed by one fundamental mode $\boldsymbol{A}_1$ that causes the observed out-of-phase oscillation [Fig.~\ref{Fig2}(a)] between the left and right droplets. This also causes the overall dBEC cloud to periodically expand and contract, and thus the frequency $\omega_{m}/(2 \pi) = 33.94$ Hz corresponds to the breathing frequency.  We remark that even at large velocity, $v > v_{\rm osc}$, the mode $\boldsymbol{A}_1$ still qualitatively explains the observed oscillation pattern although the quantitative matching between eGP and NMA improves at $v < v_{\rm osc}$. For the initial conditions, $\dot{x}_1(0) = -v_1$, $\dot{x}_2(0)=0$, $\dot{x}_3(0)=v_3$, with $v_1 \ne v_3$, the coefficients become $a_{m} = (v_3 + v_1)/(2 \omega_m)$, $a_f=\omega^2_{+}(v_3 - v_1)/(4 \omega^2_1 \omega_f \sqrt{\omega^4_1 + \omega^4_2})$, and $a_s=-\omega^2_{-}(v_3 - v_1)/(4 \omega^2_1 \omega_s \sqrt{\omega^4_1 + \omega^4_2})$. The involved oscillation frequencies in the vibrational pattern (not shown here) are given by $\omega_m/(2\pi) = 33.94 \rm Hz$, $\omega_{f}/(2\pi)= 46.85 \rm Hz$, and $\omega_{s}/(2\pi)= 19 \rm Hz$. Thus, one can systematically trigger $\boldsymbol{A}_2$ and $\boldsymbol{A}_3$ modes in addition to $\boldsymbol{A}_1$ by maneuvering the imbalance of outer droplet velocities.   Notably, we can now activate the COM motion manifested by the oscillation at $\omega_{s}/(2\pi)= 19 \rm Hz$ (which is equal to the trap frequency $\omega_x$) and the relevant mode is $\boldsymbol{A}_3$. Naturally, a larger $a_s$ indicates more vigorous COM motion that indeed takes place within the case \textbf{III} having initial condition $\dot{x}_1(0)= v$, $\dot{x}_2(0)=0$, and $\dot{x}_3(0)=0$. The coefficients can be obtained from those of the case \textbf{II} by substituting $v_1 = -v$, and $v_2 =0$. The trajectories obtained via eGP simulation are well produced by those obtained via NMA [see Fig.~\ref{Fig2}(b)] with $\omega_1/(2 \pi) = 24 \rm Hz$ and $\omega_2/(2 \pi ) = 25.7 \rm Hz $ and $\omega_m/(2\pi) \approx 33.94 \rm Hz$, $\omega_{f}/(2\pi)\approx 46.85 \rm Hz$, and $\omega_{s}/(2\pi)\approx 19 \rm Hz$, as obtained from  the case \textbf{II}. Interestingly, the coefficient $a_{m}$ vanishes for the case \textbf{IV} while the rest become $a_{f} = -v/(4 \omega_f \sqrt{\omega^4_1 + \omega^4_2})$ and $a_s = v/(4 \omega_s \sqrt{\omega^4_1 + \omega^4_2})$, rendering $x_1(t)=x_3(t)$ and $\abs{a_f} < \abs{a_s}$. Thanks to the vanishing contribution of the $\boldsymbol{A}_1$ mode, the outer droplets demonstrate the same trajectories performing in-phase oscillations and thus resulting in the strongest COM motion among the configurations considered herein. The underlying frequencies comprise of $\omega_1/(2 \pi) = 24 \rm Hz$, $\omega_2/(2 \pi) = 25.7\rm Hz$, $\omega_{s}/(2 \pi)= 19 \rm Hz$ and $\omega_{f}/(2 \pi)= 46.85 \rm Hz$. Let us remark that, by maneuvering the outer droplet velocities, one can excite the single or combination of normal modes that strongly resembles those of a classical system. We note that our analysis is rather generic, applying to arrays with large numbers of isolated droplets (see Ref.~\cite{Norcia2021}), different particle numbers or trapping frequencies. \par
{\it \textcolor{blue}{Impact of the supersolidity.}--} 
The similarity of motion between the $\rm DL_I$ state and the conventional classical spring-mass system relies on the fact that the mass of each droplet is conserved. In the SS state, however, a mass flow occurs between the droplets interconnected by the dilute superfluid,  and the analogy with classical droplet motion breaks down. This is evident from Figs.~\ref{Fig4}(a$_1$) where the case \textbf{I} is displayed. Although the central droplet remains still motionless, notably there is a particle transfer across the humps via the dilute background density, even at a low velocity such as $v = 0.4 v_{\rm osc}$[Fig.~\ref{Fig4}(a$_1$)]. A notably interesting scenario emerges for the case \textbf{IV} [Figs.~\ref{Fig4}(a$_1$)-(a$_3$)], depicted for two different velocities. The low-density peaks located left to the central droplet [Figs.~\ref{Fig1}(a$_2$)-($a_3$)] become increasingly populated during the dynamics, even though the droplets are kicked along the $+x$ direction. This implies a mass flow along the $-x$ direction via the dilute background density, revealing the existence of out-of-phase motion between the droplet arrays and background superfluid~\cite{Guo2019}. 
 Furthermore, an in-phase motion exists, which corresponds to the dipole mode and determines the COM oscillation frequency of the isolated droplets. The out-of-phase motion, however, is unique to the SS state, and its frequency is determined by the superfluid fraction.  This out-of-phase motion dcreases both the amplitude and time-period of the COM of the entire cloud in the SS state, as determined by the quantity $X_{cm} = \int x\abs{\psi(x,y,z)}^2 dx dy dz$, see Fig.~\ref{Fig1}(b). The decrease in COM mass motion fundamentally serves as an indicator of mass flow and the submergence of classicality in the vibrational motion.    \par
{\it \textcolor{blue}{Conclusions.}--} 
We analyzed the vibrational modes of a linear dipolar droplet array. A selective kicking of isolated droplets induces distinct vibrational patterns resembling those of a classical spring-mass system. At slow kick velocities,  
a normal mode analysis accurately describes the vibrational patterns. For faster kicks the droplets can touch which may induce a mass flow, so that  droplets can coalesce and separate again with different mass distributions, giving rise to different crystalline structures during the dynamics. In the case of a supersolid, classicality is eliminated and a counterflow between droplet motion and superfluid dilute background occurs. 
In the light of the recent discovery of two-dimensional supersolidity~\cite{Norcia2021,Hertkorn2021b,Poli2021,Bland2022} we expect our work to have high relevance for  assessing classicality and vibrational modes in these structures. It will also be intriguing to compare the normal mode analysis with the underlying collective excitation spectra obtained by a Bogoliubov-de Gennes approach. A finite-temperature study would be equally interesting~\cite{Sanchez2022}. 
\par 
\textit{Acknowledgements.} This work was financially supported by the Knut and Alice Wallenberg Foundation and the Swedish Research Council. Fruitful discussions with Tiziano Arnone Cardinale, Sergej Moroz are acknowledged.

\bibliographystyle{apsrev4-1.bst}
\bibliography{reference.bib}

\newpage
\pagebreak
\widetext
\clearpage

\begin{center}
  \textbf{\large Supplemental Material: Classical linear chain behavior from dipolar droplets to supersolids}\\[.2cm]
  K. Mukherjee and S.M. Reimann\\[.1cm]
  {\itshape Mathematical Physics and NanoLund, Lund University, Box 118, 22100 Lund, Sweden}\\
(Dated: \today)\\[1cm]
\end{center}
\twocolumngrid

\setcounter{equation}{0}
\setcounter{figure}{0}
\setcounter{table}{0}
\setcounter{page}{1}
\renewcommand{\theequation}{S\arabic{equation}}
\renewcommand{\thefigure}{S\arabic{figure}}
\renewcommand{\bibnumfmt}[1]{[S#1]}
\renewcommand{\citenumfont}[1]{S#1}

\section{Computational Details}

\beginsupplement
\begin{figure}
	\centering
	\includegraphics[width = 0.45\textwidth]{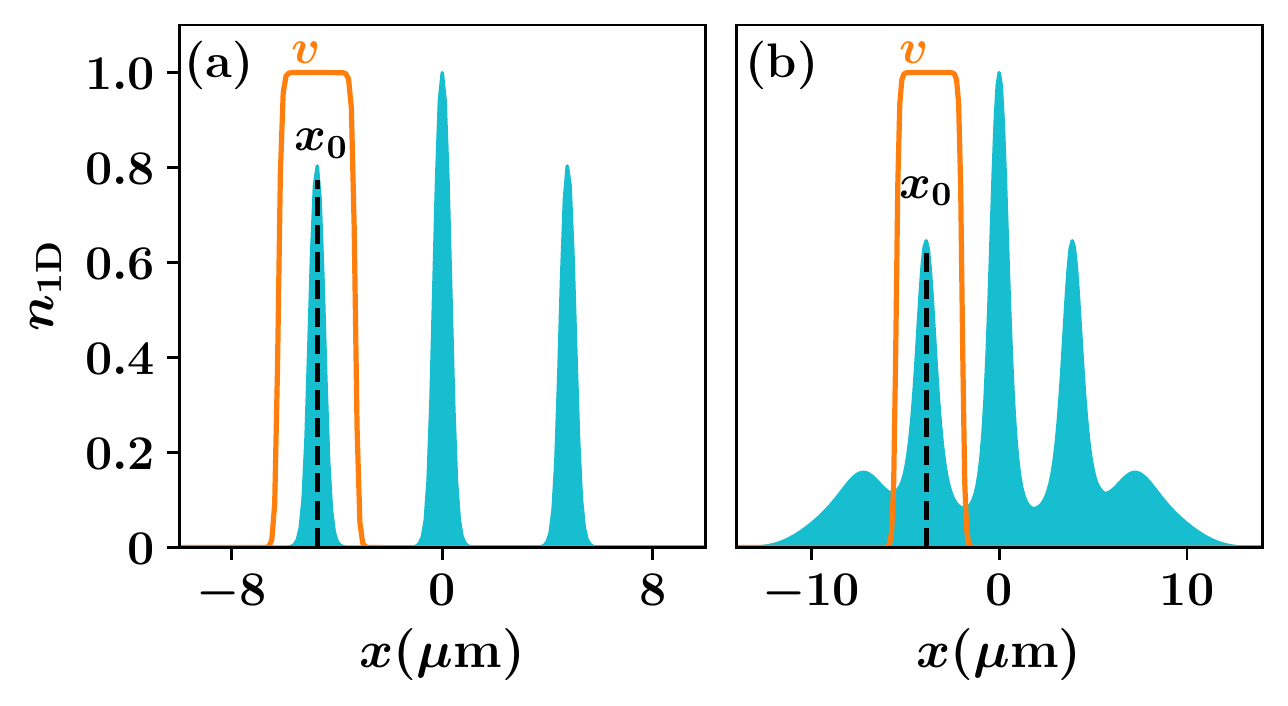}
	\caption{ Shown here are integrated density profiles (blue shaded region), $n_{\rm 1D}(z) = \int \abs{\psi(x,y,z)}^2 dx dy$, normalized with respect to the maximum $n_{\rm 1D}$, representing the (a) isolated droplets and (b) supersolid state, realized at $a_s = 84a_{0}$ and $a_s = 94_{0}$, respectively. The black dotted lines indicate the central position $(x_{0})$ of a droplet crystal. To initiate the dynamics, a droplet is kicked with a velocity, $v$, that is constant over the droplet's spatial extension; then falls swiftly but continuously to zero(see the orange curves).  }
	\label{figsupp_velo}
\end{figure}

In our numerical simulations, we cast the extended Gross-Pitaevskii equation (eGP), Eq.~\eqref{eGPE} in the main article, into a dimensionless form by rescaling the length, the time in terms of the harmonic oscillator length scale $l_{\rm osc} = \sqrt{\hbar/ m\omega_x}$, and the trap frequency $\omega_x$, respectively. The wavefunction is accordingly scaled as $\psi(\vb{r}',t) = \sqrt{l^3_{\rm osc}/N} \psi(\vb{r},t)$. Thereafter we employ split-step Crank-Nicholson method (see Ref.~\cite{crank_nicolson1947} in the main article) to solve the resulting dimensionless equation. The stationary (lowest energy) states of the dipolar Bose-Einstein condensate (dBEC) are obtained through imaginary time propagation,  effectively a  gradient descent algorithm. At each imaginary time-step of $\Delta t_i = 10^{-4}/\omega_x$ of this procedure, we apply the transformation $\frac{\psi(\vb{r}', t)}{\norm{\psi(\vb{r'}, t)}} \rightarrow 1$ . This preserves the normalization of the wavefunction, while convergence is reached as long as relative deviations of the wave function (at every grid point) and energy between consecutive time-steps are smaller than $10^{-5}$ and $10^{-7}$, respectively. This solution is taken to locate the peak densities and thus identify the central position $x_{0}$ of a droplet, see Fig.~\ref{figsupp_velo}  for the visualization. Subsequently, the wavefunction is multiplied with the phase $v \mathcal{F}(x)$, where $v$ is the velocity and $\mathcal{F}(x) = A/(A + B\cosh(L(x - x_{0})))$ is a modulating function ensuring the continuity of the wave function. We take $A = 10^{6}$, $B = 0.02$, and $L$ depends on the spatial extension of the droplet along the $x$-axis. Having multiplied the wavefunction with the above-mentioned phase we propagate the eGP equation  in real time to carry out the vibrational and collisional dynamics of dipolar quantum droplets. The simulation is performed in a 3D box characterized by a grid $(n_x \times n_y \times n_z)$ corresponding to $(512 \times 256 \times 256)$ grid points. The employed spatial discretization step is $\Delta x = \Delta y = 0.06 l_{\rm osc}$, and $\Delta z = 0.1 l_{\rm osc}$, while the time-step of the numerical integration is $\Delta t = 10^{-5}/\omega_x$. Finally, let us comment that our dynamical simulation is very well resolved upto kick velocity $v=7.5 v_{\rm osc}$ for the considered spatial discretization steps. However, for stronger kick velocity $v > 7.5 v_{\rm osc}$, since the droplets become very localized during the collision, one should reduce the spatial and temporal discretization steps further. At the same time, one should increase the number of grid points to reduce the boundary effects. The numerical simulation at this velocity scale is expected to be challenging.  
\section{Detailed Derivation of the normal modes of vibration}
As we have explicated in the main text, the vibrational motion of the three isolated quantum droplets can be modeled by resorting to a spring-mass system~\cite{Pippard}, where three masses are connected by the springs with the identical spring constant $k$ (see the Fig.~\ref{Fig0}(c) in the main text). The $x_1$, $x_2$, and $x_3$ denote the displacement of the left, center, and right masses from their equilibrium positions. In order to correctly capture the physical situation, we consider that $m_1 = m_2 = m$ and $m_2 > m$. Assuming that the spring force is linear, the equations of motion of three masses can be written as 
\begin{equation} \label{eq2}
	m\ddot{x}_1 = -k x_1 - k (x_1 - x_2)
\end{equation}
\begin{equation}\label{eq3}
	m_2 \ddot{x}_2 = -k(x_2 - x_1) - k(x_2 - x_3)
\end{equation}
\begin{equation}\label{eq4}
	m \ddot{x}_3 = -k(x_3 - x_2) -k x_3
\end{equation}
From Eq.~\eqref{eq2}, Eq.~\eqref{eq3} and Eq.~\eqref{eq4}, we get
\begin{equation}\label{eq5}
	\ddot{x}_1 = -\omega^2_1x_1 - \omega^2_1(x_1 - x_2),
\end{equation}
\begin{equation}\label{eq6}
	\ddot{x}_2 = - \omega^2_2(x_2 - x_1) - \omega^2_2(x_2 - x_3),
\end{equation}
\begin{equation}\label{eq7}
	\ddot{x}_3 = -\omega^2_1(x_3 - x_2) - \omega^2_1 x_3 ,
\end{equation}
where $\omega_1 = \sqrt{(k/m)}$, $\omega_2 = \sqrt{k/m_2}$~.
The solution of Eq.~\eqref{eq5}, Eq.~\eqref{eq6} and Eq.~\eqref{eq7} can be written in the form
$x_j = \mathcal{A}_j e^{i \omega t}$. This gives the set of equations
\begin{equation}
\mathcal{A}_1(\omega^2 - 2\omega^2_1) + \omega^2_1 \mathcal{A}_2 = 0,
\end{equation}
\begin{equation}
	\omega^2_2 \mathcal{A}_1 + \mathcal{A}_2 (\omega^2 - 2 \omega^2_2) + \omega^2_2 \mathcal{A}_3 = 0,
\end{equation}
and 
\begin{equation}
	\omega^2_1 A_2 + (\omega^2 - 2 \omega^2_1 )A_3 = 0.
\end{equation}
The above equations can be cast into a matrix of the form $\vb{M}\vb{A} = 0$, where
\begin{equation}\label{Mmatrix}
\vb{M} = 
	\begin{pmatrix}
	\omega^2 - 2\omega^2_1 & \omega^2_1 & 0 \\
	\omega^2_2 & \omega^2 - 2 \omega^2_2 & \omega^2_2
	\\
 0 & \omega^2_1	& \omega^2 - 2\omega^2_1
	\end{pmatrix}
\end{equation}
and 
\begin{equation}
	\vb{A} = \begin{pmatrix}
	\mathcal{A}_1 \\ \mathcal{A}_2 \\ \mathcal{A}_3
	\end{pmatrix}
\end{equation}
A nonzero solution exist for $\vb{A}$ only if the determinant of  $\vb{M}$ is zero. This  gives

\begin{equation}\label{chareq}
\begin{split}
&	(\omega^2 - 2\omega^2_1 )\bigg[(\omega^2 - 2 \omega^2_2)(\omega^2 - 2\omega^2_1) - \omega^2_1 \omega^2_2\bigg] - \\&
	\omega^2_1 \omega^2_2(\omega^2 - 2 \omega^2_1 ) = 0
\end{split}
\end{equation}

The roots of the Eq.~\eqref{chareq} are 
$\omega^2 = 2\omega^2_1$, and $\omega^2 = \omega^2_1 + \omega^2_2 \pm \sqrt{\omega^4_1 + \omega^4_2}$.
Plugging these values back into Eq.~\eqref{Mmatrix} we can determine the relations between $\mathcal{A}_1$, $\mathcal{A}_2$ and $\mathcal{A}_3$, which gives the three normal modes, $\mathcal{A}_1:\mathcal{A}_2:\mathcal{A}_3 = \omega^2_1: \omega^2 - 2\omega^2_1:\omega^2_1$.
For $\omega = \omega_m = \pm \sqrt{2} \omega_1$, the eigenmode is 
\begin{equation}
\vb{A}_1 = 
\begin{pmatrix}
\mathcal{A}_1 \\ \mathcal{A}_2 \\ \mathcal{A}_3
\end{pmatrix}
=
\begin{pmatrix}
1 \\ 0 \\ -1
\end{pmatrix}~.
\end{equation}
For $\omega = \omega_f = \pm [ \omega^2_1 + \omega^2_2 + \sqrt{\omega^4_1 + \omega^4_2}]^{1/2}$, the eigenmode reads
\begin{equation}
\vb{A}_2 = 
	\begin{pmatrix}
	\mathcal{A}_1 \\ \mathcal{A}_2 \\\mathcal{A}_3
	\end{pmatrix}
	=
	\begin{pmatrix}
	\omega^2_1 \\ \omega^2_{-} \\ \omega^2_1
	\end{pmatrix}.	
\end{equation}

Finally, for $\omega = \omega_s = \pm [ \omega^2_1 + \omega^2_2 - \sqrt{\omega^4_1 + \omega^4_2}]^{1/2}$, the eigenmode becomes
\begin{equation}
\vb{A}_3 = 
\begin{pmatrix}
\mathcal{A}_1 \\ \mathcal{A}_2 \\ \mathcal{A}_3
\end{pmatrix}
=
\begin{pmatrix}
\omega^2_1 \\ \omega^2_{+} \\ \omega^2_1
\end{pmatrix}.	
\end{equation}
Here, $\omega^2_{\pm} = (\omega^2_2 - \omega^2_1) \pm \sqrt{\omega^{4}_1 + \omega^{4}_2}$. 

Thus, the most general solution, $\vb{X}=(x_1, x_2, x_3)$, can be written as 
\begin{equation}
\begin{split}
	  \vb{X} = & a_{m}\vb{A}_1 \cos(\omega_m t + \phi_m) + a_{f}\vb{A}_2 \cos(\omega_f t + \phi_f) \\ & + a_{s}\vb{A}_3 \cos(\omega_s t + \phi_s)~.
\end{split}
\end{equation}

The six unknowns, $A_m$, $A_s$, $A_f$, $\phi_m$, $\phi_f$, and $\phi_s$ are determined by the six initial conditions (three positions and three velocities).
In the following for all the cases considered we have fixed initial conditions on the positions, namely, $x_1 =0$, $x_2 = 0$, and $x_3 = 0$. This gives $\phi_m = \phi_f = \phi_s = \pi/2$.
\end{document}